\documentclass[apjl]{emulateapj}
\usepackage{apjfonts}

\usepackage{graphicx}
\usepackage{amsmath}

\newcommand{\msun}{M$_{\odot}$}
\newcommand{\kms}{km~s$^{-1}$}

\newcommand{\oiii}{[O\,{\sc iii}]\,}

\begin{document}

\shorttitle{Lensed galaxies in CANDELS}
\shortauthors{Cooray et al.}

\title{CANDELS: Strong Lensing Galaxies in HST/WFC3 Imaging Data of UDS and GOODS-S}
\author{ 
Asantha Cooray\altaffilmark{1,\dag},
Hai Fu\altaffilmark{1},
Jae Calanog\altaffilmark{1},
J. L. Wardlow\altaffilmark{1},
A. Chiu\altaffilmark{1},
S. Kim\altaffilmark{1},
J. Smidt\altaffilmark{1},
V. Acquaviva\altaffilmark{2}, 
H. C. Ferguson\altaffilmark{3},
S. M. Faber\altaffilmark{4},
A. Galametz\altaffilmark{5},
N. A. Grogin\altaffilmark{3},
W. Hartley\altaffilmark{6},
D. Kocevski\altaffilmark{4},
A. Koekemoer\altaffilmark{3},
D. C. Koo\altaffilmark{4},
R. A. Lucas\altaffilmark{3},
L. Moustakas\altaffilmark{7},
J. A. Newman\altaffilmark{8}
}
\altaffiltext{\dag}{acooray@uci.edu}
\altaffiltext{1}{Department of Physics \& Astronomy, University of California, Irvine, CA 92697}
\altaffiltext{1}{Department of Physics and Astronomy, Rutgers, The State University of New Jersey, Piscataway, NJ 08854}
\altaffiltext{3}{Space Telescope Science Institute, Baltimore, MD 21218}
\altaffiltext{4}{UCO/Lick Obs. and Department of Astronomy and Astrophysics, University of California, Santa Cruz, CA 95064 USA}
\altaffiltext{5}{INAF - Osservatorio di Roma, Via Frascati 33, I-00040, Monteporzio, Italy}
\altaffiltext{6}{School of Physics and Astronomy, University of Nottingham, University Park, Nottingham NG7 2RD, UK}
\altaffiltext{7}{Jet Propulsion Laboratory, California Institute of Technology, Pasadena, CA 91109}
\altaffiltext{8}{Department of Physics and Astronomy, University of Pittsburgh, Pittsburgh, PA}



\label{firstpage}

\begin{abstract}
We present results from a search for gravitationally lensed galaxies present in  
the {\it Hubble Space Telescope (HST)}/Wide Field Camera-3 (WFC3) images of the Cosmic Assembly Near-IR Deep Extragalactic Legacy Survey (CANDELS).
We present one bona fide lens system in UDS and two compact lens candidates in the GOODS-S field. 
The lensing system in UDS involves two  background galaxies, one at $z=1.847$ lensed to an arc and a counterimage,
and the second at a photometric redshift of $z=2.32^{+0.10}_{-0.06}$ lensed to a double image.
We reconstruct the lensed sources in the source plane and find in each of the two cases the sources can be separated to a pair of galaxies.
The sources responsible for the arc are compact with effective radii of 0.3 to 0.4 kpc in WFC3 $J_{125}$-band and a total stellar mass and a 
star-formation rate of $2.1_{-0.4}^{+2.4} \times 10^7$ \msun\ and $2.3_{-1.7}^{+0.6}$ \msun\ yr$^{-1}$, respectively.
The abnormally high $H_{160}$-band flux of this source is likely due to \oiii\ emission lines with a rest-frame equivalent width about 700\AA\ for \oiii\ 
5007 \AA. The sources responsible for the double image have corresponding values of about 0.4 to 0.5 kpc, $1.4_{-0.8}^{+1.9} \times 10^9$ \msun\, 
and $8.7_{-7.0}^{+11.1}$ \msun\ yr$^{-1}$. 
Once completed CANDELS is expected to contain about 15 lensing systems and will allow statistical studies on both lensing mass profiles and 
$z \sim 2$ lensed galaxies.
\end{abstract}


\section{Introduction}

Gravitational lensing is an invaluable tool in astrophysics. It has been exploited in the past
to study properties of both the background lensed galaxies as well as the foreground lenses
\citep[see reviews by][]{Bartelmann2010,Treu2010}.  
The variety of studies enabled by lensing has
motivated searches for lensing events in various multi-wavelength astronomical datasets, from 
optical (e.g., SLACS: \citealt{Bolton2006}), sub-mm (e.g., {\it Herschel}: \citealt{Negrello2010}) to radio (CLASS: \citealt{Browne2003}).
Given that most galaxy-scale lenses involve $\lesssim 1''$ image separations, most successful searches for gravitational lenses have involved the use of high resolution imaging.
In this respect, {\it HST} surveys have been successful in uncovering 
lensed galaxies even in relatively small areas on the sky \citep{Ratnatunga1995,Barkana1999,Fassnacht2004,Moustakas2007,More11}.

The CANDELS is now obtaining WFC3 and ACS imaging data of several well-known
extragalactic fields at unprecedented depth and resolution \citep{Grogin2011,Koekemoer2011}. 
The early data include F125W, F160W (WFC3), and F814W (ACS) imaging over an area of
249 arcmin$^2$ in GOODS-S  and UDS fields. These initial imaging data are adequate for a first study of lensed galaxies.
In comparison to the work we present here \cite{More11}
identified 10 candidate lensed galaxies using ACS imaging over 0.22 deg$^2$ of the ECDFS.
Based on their success rate at identifying candidate lensing events,
we expect about 3 to 4 lensed galaxies in current CANDELS data and about 15 when completed with imaging
over 0.35 deg$^2$. Due to the ground-based and ACS data in CANDELS
fields, we do expect some fraction of the lensed galaxies to be identified already.

The paper is organized as follows: In Section~2 we present a brief summary of the candidate lens selection.
Section~3 describes surface brightness models, while in Section~4 we outline the lens models and discuss results on the three lens systems.
For lensing models we assume the best-fit concordance cosmology consistent with WMAP-7 year data \citep{Larson2011}.

\begin{deluxetable*}{lcccccccc}
\tablewidth{0pt}
\tablecaption{Properties of GOODS-S Candidate Lensed Systems
 \label{tab:detail}}
\tablehead{
\colhead{Source} & \colhead{RA$^{(a)}$} & \colhead{Dec$^{(a)}$}  & \colhead{$\mu^{(f)}$} & \colhead{$R_{\rm Einstein}$} & $\sigma_{\rm lens}^{(b)}$ & $R_e^{(g)}$ & S\'ersi
c index $n$ & $\epsilon^{(h)}$ }
\startdata
GOODS-S01 & 53.019907$^{\circ}$ & -27.770704$^{\circ}$   &              & $0.53\pm0.06$\arcsec & & $0.19 \pm 0.01$\arcsec$^{(c)}$ & $0.31 \pm 0.01$$^{(c)}$ & $0.748 \pm 0.005$$^{(c)}$\\
$z_{\rm lens}=1.22^{(i)}$     &      & & & & & & &\\
(Dark matter) &                          &                           &              &                     & $250 \pm 20$$^{(d)}$  &       &        &     $0.71_{-0.12}^{+0.09}$$^{(d)}$ \\
(lensed source)  &                     &                      & $38_{-6}^{+45}$   &             & & $0.08\pm0.01$$^{(e)}$             &  $1.2_{-0.1}^{+1.0}$$^{(e)}$    & $0.35_{-0.12}^{+0.65}$$^{(e)}$\\ 

\hline
GOODS-S02 & 53.026810$^{\circ}$ & -27.791320$^{\circ}$   &              & $0.45\pm0.07$\arcsec & &  $0.324 \pm 0.003$\arcsec$^{(c)}$              &  $4.5 \pm 0.03$$^{(c)}$               & $0.77 \pm 0.03$$^{(c)}$ \\
 $z_{\rm lens}=1.02^{(i)}$  & & & & & & & &\\
(Dark matter) &                          &                         &              &                     & $216 \pm 9$$^{(d)}$  &       &        &     $0.50_{-0.04}^{+0.27}$$^{(d)}$ \\
(lensed source) &                     &                      & $10^{+35}_{-2}$&                               &             &  $0.42_{-0.03}^{+0.07}$\arcsec$^{(e)}$ & $1.8 \pm 0.3$$^{(e)}$   &  $0.89_{-0.61}^{+0.09}$$^{(e)}$
\enddata
\tablecomments{{\it (a):} Lensing galaxy centroids from {\tt GALFIT} using F814W. 
{\it (b):} In units of km/sec. 
{\it (c):} For the $I_{814}$-band lens light profile from {\tt GALFIT} using F814W.
{\it (d):} for the lens dark matter profile from lens modeling.
{\it (e):} for the lensed $I_{814}$-band background source light profile from lens reconstruction and {\tt GALFIT} modeling.
{\it (f):} Lensing magnification.
{\it (g):} Effective radius from {\tt GALFIT} modeling.
{\it (h):} Axis ratio from {\tt GALFIT} modeling.
{\it (i):} Spectroscopic redshift from \citet{Vanzella05}.
}
\end{deluxetable*}

\begin{figure}
\epsscale{1.0}
\plotone{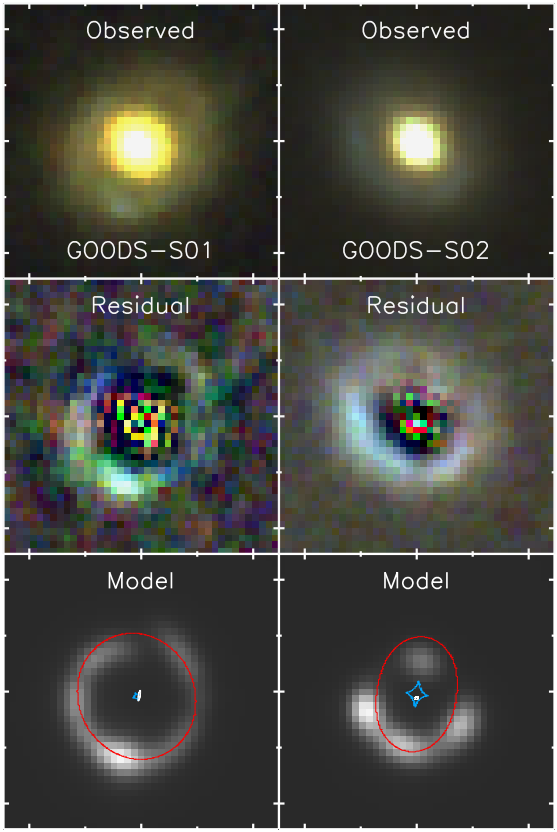}
\caption{Lens candidates in GOODS-South (GOODS-S01 and GOODS-S02). The two columns are for the two candidates and from top to bottom we show
{\it HST}/ACS $I_{814}$-band and WFC3 $H$- and $J$-band (RGB) pseudo-color images of the original and lens-subtracted residual images, and the best-fit lens model. 
The source position and shape is indicated by the white ellipse near the center of the right panels. The red and
 blue curves show the critical curve and the caustic.  Each tick mark is 1$''$.
\label{fig:goods}}
\end{figure}

\section{Lens Candidate Selection}

For the lens candidate search we make use of the CANDELS v0.5 drizzled mosaics at the pixel scale of 0.03 and 0.06$''$/pixel for WFC3 and ACS 
bands, respectively,
and involving two epochs of UDS and five epochs of GOODS-S (wide)\footnote{Mosaics are publicly available at http://candels.ucolick.org/} \citep{Koekemoer2011}. The UDS mosaic has  5$\sigma$ point source depths of 27.3, 26.3 and 25.9 (AB) magnitudes in F814W ($I_{814}$-band), F125W ($J_{125}$-band), and F160W ($H_{160}$-band), respectively. The corresponding depths for
GOODS-S mosaics are 27.7, 26.6, and 26.5 (AB). 
As an initial search for lensed galaxies in the UDS and GOODS-S fields we created individual false-color postage stamps of $\sim10^4$ potential 
lensing galaxies with $I\le26.5$~mag.
Our selection of lensed candidates is simply based on morphological information only. For this, we 
inspected individual image cutouts searching for both extended structure and color differences at the outskirts of 
galaxies that could be evidence for lensing. This process allowed 
us to make an initial selection of 20 candidate gravitationally lensed galaxies.
Through an internal poll within the CANDELS team we ranked the list of 20 and selected three
high priority targets for further studies reported in this paper.

The three lensing candidates are GOODS-S01 and GOODS-S02 (Table~1) and UDS-01 (Table~2).
GOODS-S02 was identified  by \citet{More11} as a candidate lensing system and is included as their second
highest priority lensing system from ACS images; The highest priority lensing system from \citet{More11}
is not in our candidate list as CANDELS tiles do not overlap with it.  GOODS-S01 was previously identified as a partial-ring
galaxy in GEMS and GOODS \cite{Elmegreen06}.

UDS-01 was previously identified as a lensing galaxy by \cite{Geach07} as part of a radio source followup.
 The lensed source was later identified in the CFHT Legacy Survey and was included in a snapshot {\it HST}/ACS lensing
program for strong lenses in the CFHTLS \citep{Tu09}. The CANDELS WFC3 imaging complements existing ACS data in $V_{606}$- (F606W) and $I_{814}$-band.
Furthermore this field has both CFHTLS MegaCam imaging  in the optical and UKIDSS WFCAM Ultra-Deep Survey (UDS) imaging in the near IR.
CFHTLS images reach 5$\sigma$ depths of 25.9, 26.5, 25.8, 26.0 and 24.6 magnitudes (AB), respectively, in $u$, $g$, $r$, $i$, and $z$ 
respectively with typical seeing in the UDS-01 region of about 0.8 arcsec (FWHM). We also make use of
$J$, $H$ and $K$-band images from UKIDSS UDS data release 8 (Almaini et al. in prep.), which reach $24.9$, $24.0$ and $24.6$ magnitudes 
(AB, $5\sigma$ with $2\arcsec$ aperture) respectively. 
The seeing FWHM in the region of UDS-01 range from  $0.75\arcsec$ for the $K$-band data to $0.83\arcsec$ for $H$-band.

UDS-01 involves a foreground lens galaxy which is a member of a group/cluster potential at $z=0.648$ and with a velocity dispersion of 770 km/sec
\cite{Geach07}.  
The system involves two background lensed sources with one at a measured spectroscopic redshift of 1.847 and another with a previously
estimated photometric redshift of 2.90$^{+0.18}_{-0.24}$ in \citet{Tu09} using CFHTLS and ACS $V_{606}$- and $I_{814}$-band data. 
We revise this estimate to be $2.32^{+0.10}_{-0.06}$ here with 12-band photometry involving CFHT ($u$, $g$, $r$, $i$, $z$), 
UKIDSS ($J$, $H$, $K$), ACS ($V_{606}$, $I_{814}$), and WFC3 ($J_{125}$, $H_{160}$).

\begin{deluxetable}{lcl}
\tablewidth{0pt}
\tablecaption{Properties of the UDS-01 Lensing System
 \label{tab:uds}}
\tablehead{
\colhead{Quantity} & \colhead{Value} & \colhead{Ref}}
\startdata
\multicolumn{3}{c}{Lens Galaxy}\\
RA  &  34.404790$^{\circ}$ & ACS $I_{814}$-band\\
Dec &  $-$5.224868$^{\circ}$ & ACS $I_{814}$-band\\
Redshift & 0.6459$\pm$0.0003 & Tu et al. 2009 \\
$\sigma_{\rm SIE}$ & $280.1 \pm 0.3$ \kms & lens model\\
$R_{\rm Einstein}$ & $1.183\pm0.005$\arcsec & lens model\\
 $\epsilon_d$ & $0.879 \pm 0.004$ & lens model \\
$PA_d$ & $73.5 \pm 10^{\circ}$ & lens model\\
$R_e$ & $1.170 \pm 0.004$$''$ & {\tt GALFIT} \\
      &  $8.14 \pm 0.03$ kpc &    \\
$\epsilon_s$ & $0.918 \pm 0.002$ & {\tt GALFIT} \\
$n$ (S\'ersic) &   8.0$^{(I)}$ &  {\tt GALFIT} \\
\hline
\multicolumn{3}{c}{Lens Cluster}\\
RA  &  34.389725$^{\circ}$ &  Geach et al. 2007\\
Dec & $-$5.220814$^{\circ}$ & Geach et al. 2007\\
Redshift & $0.648 \pm 0.001$ & Geach et al. 2007 \\
$\sigma_{\rm SIS}$ & $774 \pm 170$ \kms & Geach et al. 2007 \\
\hline
\multicolumn{3}{c}{Tangential Arc}\\
Redshift & $1.8470 \pm 0.0003$ & Tu et al. 2009 \\
ACS/F606W   & $22.28 \pm 0.03$ &   Photometry \\
ACS/F814W   & $22.12 \pm 0.03$ &   \\
WFC3/F125W   & $21.89 \pm 0.02$ & \\
WFC3/F160W   & $21.11 \pm 0.02$ & \\
RA$^{(II)}$ &  34.402409$^{\circ}$ & lens model\\
Dec$^{(II)}$ &  $-$5.224189$^{\circ}$ & lens model\\
$\mu$ (a)  &  $50_{-4}^{+8}$ & lens model\\
$n$(a) & $1.05 \pm 0.25$ & {\tt GALFIT}/lens reconstuction\\
$R_e$(a)  & $0.008 \pm 0.003$\arcsec & {\tt GALFIT}/lens reconstuction\\
          & $0.06 \pm 0.02$ kpc & (in F125W)\\
$\epsilon(a)$ & $0.79_{-0.38}^{+0.19}$  & {\tt GALFIT}/lens reconstuction\\
$\mu$(b)  &  $40_{-3}^{+2}$ & lens model\\
$n$(b) & $1.26 \pm 0.13$ & {\tt GALFIT}/lens reconstuction\\
$R_e$(b)  & $0.036 \pm 0.004$\arcsec & {\tt GALFIT}/lens reconstuction\\
          & $0.31 \pm 0.02$ kpc & (in F125W)\\
$\epsilon(b)$ & $0.93_{-0.32}^{+0.05}$  & {\tt GALFIT}/lens reconstuction\\
M$_{\star}$ & $2.1_{-0.4}^{+2.4}\times10^{7}$ \msun$^{(III)}$ & SED/lens magnification\\
SFR & $2.3_{-1.7}^{+0.6}$ \msun yr$^{-1}$$^{(III)}$ & SED/lens magnification\\
E(B-V) & $0.25_{-0.08}^{+0.02}$ & SED/lens magnification\\
\hline 
\multicolumn{3}{c}{Double System}\\
Redshift &$2.32_{-0.06}^{+0.10}$ & photo-z \\
ACS/F606W   & $24.06 \pm 0.04$ & Photometry  \\
ACS/F814W   & $23.72 \pm 0.04$ &   \\
WFC3/F125W   & $23.14 \pm 0.03$ & \\
WFC3/F160W   & $22.69 \pm 0.02$ & \\
RA$^{(II)}$ &  34.402245$^{\circ}$ & lens model\\
Dec$^{(II)}$ &  $-$5.223925$^{\circ}$ & lens model\\
$\mu$(a) & $2.9\pm0.2$ & lens model\\
$n$(a) & $1.35 \pm 0.25$ & {\tt GALFIT}/lens reconstuction\\
$R_e$(a)  & $0.05 \pm 0.01$\arcsec & {\tt GALFIT}/lens reconstuction\\
          & $0.5 \pm 0.1$ kpc & (in F125W)\\
$\epsilon(a)$ & $0.73_{-0.25}^{+0.23}$  & {\tt GALFIT}/lens reconstuction\\
$\mu$(b) & $3.4\pm0.1$ & lens model\\
$n$(b) & $1.50 \pm 0.25$ & {\tt GALFIT}/lens reconstuction\\
$R_e$(b)  & $0.02 \pm 0.01$\arcsec & {\tt GALFIT}/lens reconstuction\\
          & $0.2 \pm 0.08$ kpc & (in F125W)\\
$\epsilon(b)$ & $0.79_{-0.38}^{+0.19}$  & {\tt GALFIT}/lens reconstuction\\
M$_{\star}$ & $1.4_{-0.8}^{+1.0}\times10^{9}$ \msun$^{(III)}$ & SED/lens magnification\\
SFR & $8.7_{-7.0}^{+11.1}$ \msun $yr^{-1}$$^{(III)}$ & SED/lens magnification\\
E(B-V) & $0.29_{-0.12}^{+0.08}$ & SED/lens magnification
\enddata
\tablecomments{
{\it (I):} {\tt GALFIT} models converge when $n>4$.
{\it (II):} Position of the best-fit brightest source component from lens modeling using F125W (WFC3 $J_{125}$-band) image. 
{\it (III):}  Corrected for lensing magnification.
}
\end{deluxetable}

\begin{figure*}
\epsscale{1.0}
\plotone{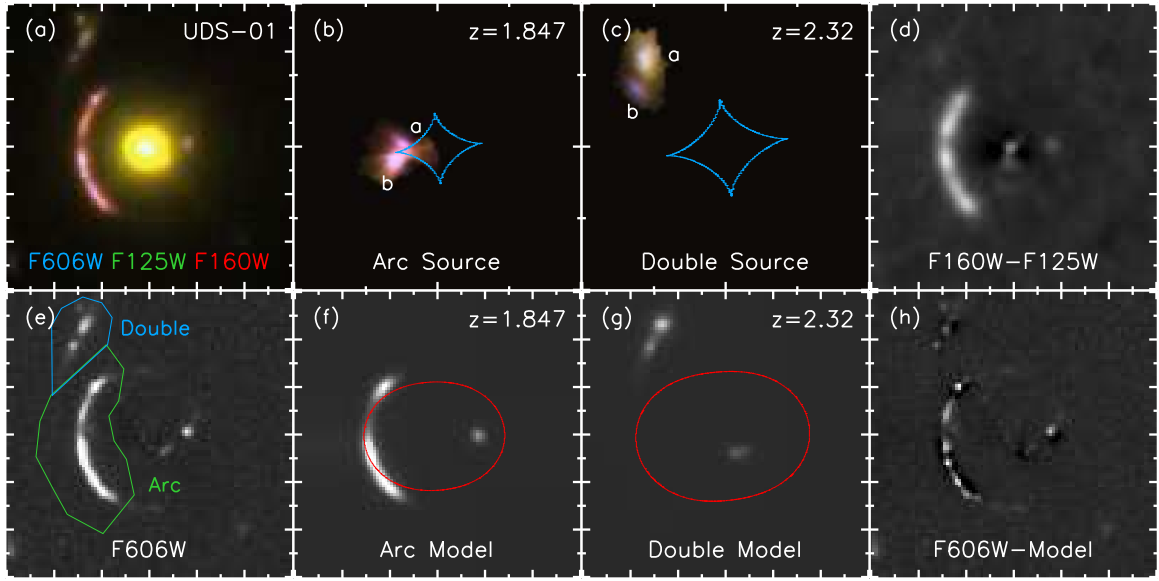}
\caption{Double source plane system UDS-01. {\it (a):} {\it HST}/ACS $I_{814}$- and WFC3 $J_{125}$ and $H_{160}$-band (RGB) pseudo-color image.
{\it (b)} and {\it (c):} the lens reconstructed pseudo-color images of the sources responsible for the arc and double lensed images, respectively.
The blue curve is the caustic at the respective source plane.
We label the two components that are lensed to produce each of the arc and the double with 'a' and 'b';
their properties are listed in Table~2. 
The corresponding lens models for $V_{606}$-band observations are shown in panels {\it (f)} and {\it (g)}, where the
red curve is now the critical line in the image plane.
These best-fit models can be compared directly to lens-substracted F606W image in panel {\it (e)}.
The difference between observed images and lensing models is shown in panel {\it (h)}.
{\it (d):} The difference between $J_{125}$ scaled to $H_{160}$ using the SED and the observed $H_{160}$.
We believe this excess originates from strong \oiii 4959 an 5007 emission that falls in F160W filter for a source at $z=1.847$.
In panel {\it (e)} we show the irregular apertures  used for photometry of the arc and north-west images of the double source
to construct SEDs shown in Fig.~3. 
Each tick mark is 1$''$; note that the reconstructed source images in {\it (b)} and {\it (c)} are zoomed-in relative to other panels. 
Through lensing magnification, we have achieved a spatial resolution of order 0.01$''$ or 100 pc in the source plane.
\label{fig:uds}}
\end{figure*}

\begin{figure*}
\epsscale{1.20}
\plotone{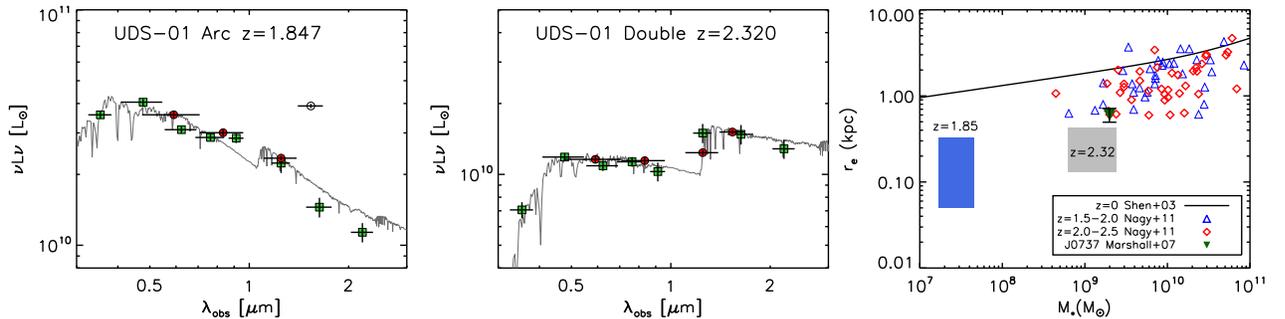}
\caption{Spectral energy distribution (SED) of the arc {\it (left)} and north-west images of the double {\it (middle)}.
The luminosity scale is magnification corrected. The photomtery points are from {\it HST} (circles) and CFHT and UKIDSS (squares). 
The curves show the best-fit SED using  \citet{Bruzual03} templates. 
{\it Right:} The size-stellar mass relation of $1.5 < z < 2.5$ galaxies. The two boxes mark the allowed 68\% confidence range of the sources that
produce the arc and the double source. The open triangles show the measurements at $1.5 < z <2.5$ by \citet{Nagy11} using WFC3 F160W.  
The filled triangle
is a previous result on a lensing galaxy at $z=0.58$ \citep{Marshall07}. The solid line is the $z=0$ relation from SDSS \citep{Shen03}. 
We use WFC3 F125W here for the size measurement as F160W flux has \oiii emission lines.
\label{fig:sed}
}
\end{figure*}

\section{Surface Brightness Models}

The two-dimensional surface brightness profile of each of the three foreground lensing galaxies were modeled using the 
{\tt GALFIT} 3.0 routine \citep{Peng10}  in order to perform 
the foreground lens-background source separation shown on Figs.~1 and 2. A 12$\times$12$''$ region was cut out from each of the
UDS and GOODS-S mosaics and was used as the input image 
for galaxy profile modeling. 

For {\tt GALFIT} modeling the foreground  lensing galaxies were assumed to have a {\it S\'ersic} profile of the form 
\begin{equation}
\Sigma(R) = \Sigma_{e}\exp[-b_{n}((R/R_{e})^{1/n}-1)]\, ,
\end{equation}
where $R_e$ is the scale radius and $n$ is the index.
Since the only object of interest for {\tt GALFIT} modeling was the lensing galaxy for each image, all other sources including the lensed sources were masked
for this purpose. 

To improve the accuracy  and decrease the level of degeneracy of the morphological parameters such as the integrated magnitude, {\it S\'ersic} index,  and effective radius, PSFs for WFC3 data generated internally by CANDELS \citep{Koekemoer2011}
were used as an input for convolution in the modeling. {\it HST}/WFC3 noise rms maps were also utilized instead of {\tt GALFIT}'s internal sigma
image generating algorithm to improve the uncertainties reported in the output parameters. 
The residual after accounting for the central lensing galaxy profile is taken to be the lensed flux of the background source. These 
residual images are used directly in lens modeling.

\section{Lens Models and Results}

We use the publicly available {\tt Lenstool}
 software \citep{Jullo07} in combination with the IDL routine {\tt AMOEBA\_SA} to optimize the lens model and the shape of the lensed galaxy. 

For the candidate lenses GOODS-S01 and GOODS-S02, we assume that each
of the foreground lens galaxies is embedded in a dark matter halo that can be described as an elliptical singular isothermal sphere (SIS). The halos are fixed to the centroids of the galaxies as determined from {\tt GALFIT}. For the lensed galaxies, we assume exponential disk profiles with
$\Sigma(r)=\Sigma_0 \exp{(-r/r_s)}$. There are a total of nine free parameters --- the position, the intrinsic magnitude, the scale length, the ellipticity, and the position angle (PA) of the source ($x$, $y$, $m_s$, $r_s$, $\epsilon_s = 1-b/a$, and $\phi_s$), and the ellipticity, the PA, and the velocity dispersion of the dark matter halo ($\epsilon_d$, $\phi_d$, and $\sigma_{\rm SIS}$). 

UDS-01 is a more complicated case. The central lensing galaxy at $z = 0.6459$ belongs to an X-ray detected galaxy group at $z = 0.648$. Hence, besides the elliptical SIS halo centered on the lensing galaxy, we also include a circular SIS halo at the center of the galaxy group 56\arcsec\ NW of UDS-01 with its velocity dispersion fixed to that of the member galaxies \citep[774 \kms\ at $z = 0.648$][]{Geach07}. In addition, there are two \emph{pairs} of background galaxies that are strongly lensed by the joint gravitational potential of the central galaxy and the group: one at $z = 1.847$ corresponding to the tangential arc, the other at a photometric redshift of $z \sim 2.32$ 
corresponding to the double system further out (see Fig.~\ref{fig:uds}). Because the tangential system provides better constraints to the dark matter distribution, we use the tangential system to find the parameters of the SIS halo associated with the central lensing galaxy, then we use this solution to constrain the positions and shapes of the double system. Again we assume exponential profiles for the background galaxies.

Our fitting procedure is as follows. For an initial set of parameters describing the source and the deflector, we use {\tt Lenstool} to generate a deflected image of the source. We then convolve it with the WFC3 PSFs at each band and subtract it from the observed lens-subtracted image to compute the $\chi^2$ value. This process is iterated with AMOEBA\_SA to find the parameters that minimize the $\chi^2$ value. AMOEBA\_SA is based on the IDL multidimensional minimization routine AMOEBA with simulated annealing added (E.~Rosolowsky, private communication). We adopt an initial ``temperature" of 100 and decrease it by 40\% in each subsequent calls of AMOEBA\_SA.

\subsection{GOODS-S01 and GOODS-S02}

Fig.~1 shows the best-fit lensing models for the Einstein ring candidate GOODS-S01 and the cusp lens GOODS-S02
and Table~1 summarizes lens model properties.
For lens modeling, we assume that the background galaxies in both systems are at $z=2$. We attempted to extract the photometric redshift of the background galaxies by SED fitting to CANDELS and archival residual fluxes. 
Both systems, however, are compact with each having an Einstein radius of $\sim$ 0.5$''$ and an accurate separation of the 
lens profile from the background source flux was challenging. 
The reduced-$\chi^2$ value for the best-fit lens model is 1.1 and 1.5 for GOODS-S01 and -S02, respectively.

The dark matter profile of GOODS-S02 lens has an ellipticity of $0.50_{-0.04}^{+0.27}$, while the
the Einstein radius is $0.45\pm0.07$$''$. These are consistent with values of
0.48 and  0.4\arcsec reported in \citet{More11}, respectively.
The lens galaxies have effective radii of $0.64 \pm 0.08$ kpc (GOODS-S01) and  $0.38 \pm 0.04$ kpc (GOODS-S02). 
Using the best-fit velocity dispersion and axis ratios, we estimate enclosed masses within critical (Einstein) radii
of $(4.8 \pm 1.8) \times 10^{11}$ M$_{\sun}$ and $(3.8 \pm 1.7) \times 10^{11}$ M$_{\sun}$ for GOODS-S01 and S02, respectively. 
The corresponding $I_{814}$-band luminosities for apertures matched to the critical radii of the two galaxies
are $(7.1  \pm 0.3) \times 10^{10}$ L$_{\sun}$ and $(1.1  \pm 0.3) \times 10^{11}$ L$_{\sun}$.
We find total mass-to-light ratios of $6.7\pm1.1$ and $3.5 \pm 1.6$ out to critical radii of GOODS-S01 and S02, respectively. 
Note that there is an additional systematic error of $\pm 1.5$ to 2.0 in the $M/L$ ratio associated with the 
unknown background source redshifts for the two lens systems.

\subsection{UDS-01}

Fig.~2 shows the best-fit lens model and the lensed galaxy image reconstruction directly in the source plane.
Table~2 summarizes lens modeling, profile fitting, and SED modeling results.
To measure source properties we make use of WFC3/F125W image.
The arc and counterimage, taking the form of a cusp lens, involve an extend source with at least two
peaks of emission.  The source responsible for double image is also made up of two separate galaxies in the source plane.  
In Table~2 we separate the properties of each of these galaxy pairs with labels $(a)$ and $(b)$, while the two components
are identified in Fig.~2(b) and (c) panels.

In Fig.~3 we show the SEDs of the sources responsible for the lensed arc and the north-west double source, with the luminosity scale corrected
for the magnification  In making these SED plots, instead of using {\tt GALFIT} to remove the lens galaxy, we take advantage of the asymmetry 
of the lensed system and create residual
images by subtracting a mirrored image from the original image. This removes the need for accurate galaxy profiles for the lens.
Irregular-sized apertures are made to enclose most of the
flux from the arc and the NW-double in the CFHT and UKIDSS images as highlighted in Fig.~2(e). 
The WFC3/F160W photometry of the arc is abnormally high.
This is  probably due to the redshifted \oiii 4959 and 5007 lines, as both fall within the WFC F160W band for a source at $z=1.847$. 
This hypothesis is supported by the fact that the bandpass in WFC3 begins at 1.4 $\mu$m, while UKIRT WFCAM $H$-band
filter used in UKIDSS has effectively a zero throughput at $\lambda < 1.45$ $\mu$m and misses the \oiii emission lines.

Stellar population synthesis models are built for exponentially declining star formation histories with SFR(t) $\propto$ exp($-t/\tau$) 
with various $\tau$ and ages \citep{Bruzual03},
with the restriction that template ages are less than the age of the universe at the corresponding redshift. For each 
template we fit for stellar mass and intrinsic extinction to match the observed SEDs, assuming the extinction law of 
\citet{Calzetti94} and parameterized by $E(B-V)$.  The templates that gives the minimum $\chi^2$ values are chosen 
as the best fit.  Since the SED templates do not include emission lines, we did not include the F160W data point for $z=1.847$ source 
in the modeling. The quoted best-fit stellar masses and SFRs are corrected for magnification using amplification factors calculated for the 
photometry apertures using the best-fit lensing model (Table~2).

We measure the difference of SED predicted flux scaled from F125W filter to F160W and the observed F160W 
flux as a measurement of the \oiii intensity. Fig.~2(d) shows that this difference is not localized to a specific region on the arc
suggesting that the \oiii emission is spatially distributed broadly across the two components responsible for the arc.
Assuming \oiii 4959 to 5007 ratio is 0.33, we estimate $F_\lambda(5007) \sim 8.9 \times 10^{-19}$ergs s$^{-1}$ cm$^{-2}$ \AA$^{-1}$
and a rest-frame equivalent width for the \oiii emission of $\sim$ 700\AA. 
The suggestion that the arc has high \oiii emission is consistent with the discovery of star-bursting
dwarf galaxies in CANDELS with strong \oiii emission lines falling in the WFC3 $J_{125}$-band \citep{vanderwel11}.
Our source is at a slightly higher redshift with \oiii in the WFC3 $H_{160}$-band.
The \citet{vanderwel11} sources have rest-frame equivalent widths as high as 1000\AA\  and \oiii fluxes a factor of 3 to 4
higher than that of the arc after correcting for magnification. The stellar mass and the SFR we find the arc source with SED fits
are consistent with the suggestion that the two sources are dwarf galaxies with a star-burst phase with $M_\star/SFR$ of around 9 to 10 Myr.

Based on {\tt GALFIT} modeling of the reconstructed source profile, the sources responsible for the arc are compact 
with effective radii around 0.05 to 0.5 kpc,   while the two sources responsible for the double images have effective radii of 0.2 to 0.6 kpc. 
In Fig.~3 right panel we summarizes these sources in the stellar mass-size and compare to other samples of galaxies.
As summarized there the arc sources are not only dwarf galaxies, or smaller than typical ``dwarf'' galaxies with a few times 10$^7$ M$_{\sun}$
stellar mass, the galaxies are bursting with stars and have high \oiii emission. The lensing system is radio bright \citep{Geach07}
and thus a high resolution map of the radio emission will be useful to have to narrow down the source of \oiii emission, especially
if either or both of the two compact, star-bursting galaxies magnified to the arc host an AGN.

To summarize this {\it Letter}, we have presented results from a search for lensed galaxies in early CANDELS imaging data involving two candidates
in GOODS-South and a bona fide lensing system in the UDS field. 
We reconstruct the lensed sources in the UDS system and through {\tt GALFIT} and SED model fits to the total
aperture fluxes establish their properties including effective radius, stellar mass, star-formation rate, among others.
The pair of galaxies lensed to the arc and a counterimage show bright \oiii line emission in the WFC $H_{160}$-band. The sources are compact and 
are undergoing a star-bursting phase. They are some of the smallest galaxies with lowest stellar masses at $z > 1.5$ and falls
below in size and stellar mass when compared to even dwarf galaxies at these redshifts.

\acknowledgments
Financial support for this work was provided by NASA
through grant HST-GO-12060 from
the Space Telescope Science Institute, which is operated
by Associated Universities for Research in Astronomy,
Inc., under NASA contract NAS 5-26555.
Partly based on observations obtained with CFHT/MegaCam.
The UKIDSS project is defined in \cite{Lawrence07}.





\end{document}